\documentclass{revtex4}

\def\bea{\begin{eqnarray}}
\def\eea{\end{eqnarray}}
\def\bec{\begin{center}}
\def\ec{\end{center}}

\def\beq{\begin{equation}}
\def\eeq{\end{equation}}

\def\f{\frac}

\def\f#1#2{\frac{#1}{#2}}

\def\pr{\prime}
\def\l{\left}
\def\r{\right}

\begin{document}
\draft
\tighten
\preprint{KAIST-TH 03/??}
\title{\large \bf A QCD Axion from Higher Dimensional Gauge Field}
\author{Kiwoon Choi\footnote{kchoi@hep.kaist.ac.kr}}
\address{Department of Physics, Korea Advanced Institute of Science and Technology\\ Daejeon
305-701, Korea}
\date{\today}
\begin{abstract}
We point out that a QCD axion solving the strong
CP problem can arise naturally  from 
parity-odd gauge field $C_M$
in 5-dimensional (5D) orbifold field theory.
The required axion coupling to 
the QCD anomaly comes from 
the 5D Chern-Simons coupling, and   
all other unwanted $U(1)_{PQ}$ breaking axion couplings can be 
avoided naturally by the 5D gauge symmetry of $C_M$ and
the 5D locality.
If the fifth dimension is warped, the resulting
axion scale is suppressed by small warp factor
compared to the Planck scale, 
thereby the model can generate naturally
an intermediate axion scale $f_a=10^{10}\sim 10^{12}$ GeV.

\end{abstract}
\maketitle


The strong CP problem  is a naturalness
problem associated with
that CP is conserved  by the strong interactions but
not by the weak interactions \cite{strongcp}.
If the $\eta^\prime$ meson receives a mass of order $\Lambda_{QCD}$
in a manner consistent with the anomalous $U(1)_A$ Ward identity,
one can not avoid CP-violaton associated with
the phase angle
$\bar{\theta}=\theta_{QCD}+{\rm Arg}{\rm Det}(M_uM_d)$
where $\theta_{QCD}$ is the bare QCD vacuum angle and $M_{u,d}$
denotes the $3\times 3$ mass matrices of the up and down-type quarks.
The observed CP violations in $K$ and $B$-meson
system suggest that $M_{u,d}$ are complex 
with phases of order unity, yielding the Kobayashi-Maskawa phase
$\delta_{KM}\approx 1$.
On the other hand,  the non-observation
of the neutron electric dipole moment implies
$({m_u}/{m_d})\,|\bar{\theta}|\,\lesssim\, 10^{-10}$
for the up to down quark mass ratio 
$m_u/m_d$.
This raises a question why the 
phase combination $\bar{\theta}$ is so small compared to
the other phase combination $\delta_{KM}$.

\medskip

There are presently three possible solutions to the strong
CP problem. One simple solution is that the up quark is massless,
rendering all CP violations associated with $\bar{\theta}$ vanish.
A massless up quark is not necessarily in conflict with the
results of chiral perturbation theory since an 
effective up quark mass can be mimicked by instanton effects
\cite{m_u=0}.
A second solution is to assume that CP is an exact symmetry
of the underlying high energy theory, but is spontaneously
broken in a specific manner to yield $|\bar{\theta}|\lesssim 10^{-10}$
\cite{nelsonbarr}.
The third, perhaps the most popular, solution is to introduce
a global $U(1)_{PQ}$ symmetry which is explicitly
broken by the QCD anomaly \cite{pq}.
At low energies, $U(1)_{PQ}$ is non-linearly
realized, leading to
a light pseudo Goldstone boson \cite{axion}, the axion, whose decay
constant $f_a$ is constrained as
$10^{10} \,\, {\rm GeV} \lesssim
f_a \,\, \lesssim 10^{12}\,\, {\rm GeV}$
 by astrophysical and cosmological 
considerations \cite{strongcp}.

\medskip

In regard to the axion solution,
there are two theoretical problems  to be understood. One is
how to suppress the unwanted explicit breakings of $U(1)_{PQ}$ 
which would spoil the Peccei-Quinn mechanism
for dynamical relaxation of $\bar{\theta}$
\cite{quantumgravity}.
This problem becomes particularly serious  when quantum gravity
effects are taken into consideration.
A commonly adopted idea for this problem is
that there exist additional continuous and/or discrete gauge symmetries
which forbid all unwanted $U(1)_{PQ}$ breakings. 
However if implemented within 4-dimensional (4D) effective field theory,
this idea appears to be rather contrived,
and the introduced gauge symmetries do not have any simple
connection to the resulting axion.
The second problem  is how to get naturally 
an intermediate axion scale $f_a = 10^{10}\sim
10^{12}$ GeV, while incorporating
the Planck scale $M_{Pl}\approx 10^{18}$ GeV and/or 
the grand unification scale $M_{GUT}\approx 10^{16}$
GeV. 
In this paper, we wish to point out that the enough suppression
of unwanted $U(1)_{PQ}$ breakings and also an intermediate
axion scale
can be accomplished naturally in 5D orbifold field theories
if the axion originates from a higher dimensional 
parity-odd gauge field $C_M$.
As we will see, such 5D theories provide a natural
framework in which the 5D gauge symmetry of $C_M$ and the 5D locality
assure that all unwanted explicit breakings of $U(1)_{PQ}$ other than
the QCD anomaly are suppressed enough not to spoil the axion
solution to the strong CP problem.
Also if the fifth dimension is warped \cite{rs},
the resulting axion decay constant is suppressed by
an exponentially small warp factor compared to $M_{Pl}$,
thereby the model can generate naturally an intermediate axion scale.

\medskip

Let us briefly discuss how much  should the 
unwanted $U(1)_{PQ}$ breakings be suppressed to keep
the axion solution to work.
If $U(1)_{PQ}$ is nonlinearly realized,
one can always choose a field basis for which only the
axion transforms under $U(1)_{PQ}$ as
$a \rightarrow a+{\rm constant}$,
while all other fields are invariant.
Note that in this field basis, {\it non-derivative} axion couplings
represent explicit $U(1)_{PQ}$ breaking.
Then the axion effective action at low energy scales can be written as
\beq
{\cal L}_{\rm axion}=
\f{1}{2}\partial_\mu a\partial^\mu a 
+\frac{1}{32\pi^2}\l(\f{a}{f_a}+
\bar{\theta}\r)\l(F\tilde{F}\r)_{QCD}
+V_{\rm HE}(a/f_a)\,,
\eeq
where $F\tilde{F}=\f{1}{2}\epsilon^{\mu\nu\rho\sigma}
F^a_{\mu\nu}F^a_{\rho\sigma}$ for the gauge field
strength $F^a_{\mu\nu}$
and we have ignored $U(1)_{PQ}$ invariant
derivative couplings of axion.
Here there are two types of $U(1)_{PQ}$ breakings: 
the axion coupling to the QCD anomaly
$(F\tilde{F})_{QCD}$ and a high energy axion potential $V_{\rm HE}$
induced by {\it all other} $U(1)_{PQ}$ breaking
effects  such as quantum gravity effects.
After integrating the QCD degrees of freedom, $(F\tilde{F})_{QCD}$ yields 
a low energy axion potential \cite{strongcp}
\beq
\label{leaxion}
V_{\rm LE}= -f_\pi^2m_\pi^2\sqrt{\f{m_u^2+m_d^2+2m_um_d
\cos (a/f_a+\bar{\theta})}{(m_u+m_d)^2}}\,,
\eeq
where $f_{\pi}\approx 94$ MeV is the pion decay constant
and $m_\pi\approx 135$ MeV is the pion mass.
The conventional axion solution to the strong CP
problem relies upon an assumption that $U(1)_{PQ}$ breakings
other than the QCD anomaly are all suppressed enough to have
\beq
\label{heaxion}
V_{\rm HE} \lesssim 10^{-10} f_\pi^2 m_\pi^2\,.
\eeq
If this assumption holds true, the axion 
vacuum expectation value is determined by $V_{\rm LE}$ 
to cancel $\bar{\theta}$, i.e. $|\langle a\rangle/f_a+
\bar{\theta}|\lesssim 10^{-10}$, thereby
solves the strong CP problem.
If not, the resulting QCD would not conserve
CP in general.
The problem is that global symmetries are broken
generically by quantum gravity effects,
so it is highly nontrivial to tune
high energy dynamics to preserve $U(1)_{PQ}$ to
an accuracy  satisfying (\ref{heaxion})
\cite{quantumgravity}.
Therefore, to be a true solution for the strong CP problem,
one needs to provide a rationale for why
all explicit breakings of $U(1)_{PQ}$ other than the QCD anomaly
are so suppressed.
In the following, we will see that 
this can be accomplished  naturally in 5D orbifold field theory
when the corresponding axion originates from
a parity-odd 5D gauge field.
\medskip

To proceed, let us consider a 5D orbifold field theory 
compactified on $S^1/Z_2$
with 5D metric $ds^2=G_{MN}dx^Mdx^N=\eta_{\mu\nu}dx^\mu dx^\nu
+dy^2$ ($M,N=0,1,2,3,5$; $\mu,\nu=0,1,2,3$),
where the fundamental
domain of $S^1/Z_2$ is represented by $0\leq y\leq \pi R$. 
The model contains a $Z_2$-odd 5D $U(1)$ gauge field $C_M$,
$Z_2$-even 5D gauge fields $A^a_M$,
generic bulk matter fields $\Phi$ and brane mater fields $\phi$,
obeying the following boundary conditions:
\bea
&& C_\mu(-y)=-C_\mu(y), \quad
C_5(-y)=C_5(y),\quad C_M(y+2\pi R)=C_M(y)\,,
\nonumber \\
&& A^a_\mu(-y)=A^a_\mu(y),\quad
A^a_5(-y)=-A^a_5(y),\quad
A_M^a(y+2\pi R)=A^a_M(y)\,,
\nonumber \\
&& \Phi(-y)=\pm \Phi(y), \quad \Phi(y+2\pi R)=\pm \Phi(y).
\nonumber 
\eea
Then the model possesses a $U(1)_C$ gauge symmetry under which
\bea 
\label{cgauge}
C_M\rightarrow C_M+\partial_M\Lambda\,,
\quad
\Phi\rightarrow e^{-iq_{_\Phi}\epsilon (y)\Lambda}\Phi\,,
\quad \phi \rightarrow \phi\,,
\eea
where $q_{_\Phi}$ is the $U(1)_C$ charge of
$\Phi$ and $\epsilon (y)=-\epsilon (-y)=\epsilon (y+2\pi R)=1$.
Here the gauge function $\Lambda(x,y)$ 
satisfies 
$\Lambda(-y)=-\Lambda(y)$ and $\Lambda(y+2\pi R)=\Lambda(y)$, so
$$
\Lambda(y=0)=\Lambda(y=\pi R)=0\,,
$$
and the gauge covariant derivatives of $\Phi$ and $\phi$ are given by
$D_M\Phi=\l(\nabla_M+iq_{_\Phi}\epsilon (y) C_M\r)\Phi$
and
$D_M\phi=\nabla_M\phi$
where the $Z_2$-even gauge fields $A^a_M$ are contained in $\nabla_M$. 
Note that all brane fields are neutral under $U(1)_C$, and
$\Phi$ and $D_M\Phi$ have the same $U(1)_C$ transformation since 
$\Lambda(y)\frac{d}{dy}\epsilon (y)=2\Lambda(y)\l(\delta(y)
-\delta (y-\pi R)\r)=0$.
Throughout this paper, we will assume that the standard model (SM)
gauge fields originate from $A^a_M$.

\medskip

The 4D scalar field $C_5$ behaves like a pseudo-Goldstone
boson. This feature has been used recently to get an inflaton \cite{inflaton},
or the SM Higgs boson \cite{higgs}, or even a quintessence
\cite{pilo},
from higher dimensional gauge fields.
Here we wish to explore the possibility that $C_5$ corresponds to a
QCD axion solving the strong CP problem.
Basically we will examine the possible explicit breakings of 
$U(1)_{PQ}$ for
\beq
\label{pq}
U(1)_{PQ}: \quad C_5\rightarrow C_5+\mbox{constant}.
\eeq
The action of our 5D theory can be written as
\bea
S_{5D}=\int dy\,d^4x \,\l[
\,{\cal L}_B+\delta(y)\,{\cal L}_b+
\delta(y-\pi R)\,{\cal L}^{\prime}_b\,\r]\,,
\eea
where ${\cal L}_B$ and ${\cal L}_b$, ${\cal L}^\pr_b$ denote
the local bulk and brane lagrangian densities, respectively.
The $U(1)_C$ gauge symmetry ensures that
$C_M$ appears in the bulk lagrangian ${\cal L}_B$ 
{\it only} through either 
$C_{MN}=\partial_M C_N-\partial_NC_M$, or 
$D_M\Phi=(\nabla_M+iq_{_\Phi}\epsilon(y)C_M)\Phi$, or a term of the form  
\bea
\label{anomaly}
\Delta{\cal L}_B=
C_M\Omega^M =C_M\partial_N\omega^{MN}\,,
\eea
where $\Omega^M$ is {\it divergenceless}, so 
can be written as $\Omega^M=\partial_N\omega^{MN}$
for a two-form valued {\it local} functional 
$\omega^{MN}=-\omega^{NM}$.
Note that $\Omega^{N}$ is required to be gauge invariant,
however $\omega^{MN}$ itself does not have to be gauge invariant.
Typical examples of divergenceless $\Omega^N$ 
with gauge {\it non-invariant} $\omega^{MN}$ would be
$\epsilon^{MNPQR}F^a_{NP}F^a_{QR}$ and 
$\epsilon^{MNPQR}R_{NP}R_{QR}$ for which
$C_M\Omega^M$ correspond to the well-known Chern-Simons (CS)
couplings of $C_M$ to the Yang-Mills field strength
$F^a_{MN}$ (of $A^a_M$) and the Riemann curvature two form $R_{MN}$.
As for the brane lagrangians, since 
all bulk and brane matter fields are $U(1)_C$ invariant 
at $y=0$ and $\pi R$, while $C_5$ transforms as
$C_5(0,\pi R)\rightarrow C_5(0,\pi R)+\partial_5\Lambda(0,\pi R)$,
$C_5$ can appear in ${\cal L}_b$ and ${\cal L}^\prime_b$
{\it only} through 
$C_{\mu 5}=\partial_\mu C_5-\partial_5 C_\mu$.

\medskip
As can be easily recognized,
$U(1)_C$ and the 5D locality severely constrain
the possible nonderivative couplings of $C_5$,
and therefore $U(1)_{PQ}$ breakings, in $S_{5D}$.
In the field basis in which $U(1)_C$ and $U(1)_{PQ}$ are
realized as (\ref{cgauge}) and (\ref{pq}), 
$U(1)_{PQ}$  is broken {\it only} by terms involving
$D_5\Phi=(\nabla_5+q_{_\Phi}\epsilon(y)C_5)\Phi$  or terms of the form
(\ref{anomaly}).
If all bulk matter fields are neutral under $U(1)_C$,
i.e. all $q_{_\Phi}=0$, which is 
stable against quantum gravity effects,
$U(1)_{PQ}$ would be broken only by terms
of the form (\ref{anomaly}).
In such case,
under a $U(1)_{PQ}$ transformation
$\delta C_5=\alpha=$ constant, we have
\bea
\label{pqbreaking}
\delta_{PQ}S_{5D}=\alpha \int dy \int d^4x \,\partial_\mu\omega^{5\mu}\,,
\eea
thus {\it $U(1)_{PQ}$ is broken only by 
4D total divergence}.
Symmetry breaking by 4D total  divergence $\partial_\mu\omega^{5\mu}$
is dominated absolutely by the {\it gauge non-invariant} piece of
$\omega^{5\mu}$, particularly by Yang-Mills instantons \cite{thooft},
since the gauge invariant piece of $\omega^{5\mu}$ vanishes
rapidly at $|x|\rightarrow \infty$.
We thus  consider
\bea
\partial_\mu\omega^{5\mu}=\kappa_h\l(F\tilde{F}\r)_{\rm hidden}+
\kappa_s\l(F\tilde{F}\r)_{QCD}+\kappa_w\l(F\tilde{F}\r)_{\rm weak}+...
\eea
where the ellipsis stands for {\it other} 4D total divergences,
e.g.  $\epsilon^{\mu\nu\rho\sigma}R_{\mu\nu}R_{\rho\sigma}$
or $\partial_\mu J^\mu$ for a gauge invariant current $J^\mu$,
which give a either null or negligible
contribution to $V_{\rm HE}$ satisfying the bound (\ref{heaxion}).
To be general, here
we have included  the instanton number of
hidden gauge fields confining at $\Lambda_h\gg\Lambda_{QCD}$,
as well as the instanton numbers of
the QCD and electroweak gauge fields.
Obviously the above breakings of $U(1)_{PQ}$ by Yang-Mills instantons
arise from the CS couplings of $C_M$ in the 5D action:
$$\int \Delta {\cal L}_B=
\int \kappa \,C\wedge F\wedge F.
$$
To solve the strong CP problem, one needs the low energy
axion potential (\ref{leaxion}) induced by 
$(F\tilde{F})_{QCD}$, so 
needs $\kappa_s\neq 0$.
$(F\tilde{F})_{\rm weak}$ would induce a high energy axion potential
whose size is highly model-dependent \cite{quintaxion_choi},
however the resulting $V_{\rm HE}\ll 10^{-10}f_\pi^2m_\pi^2$, so not harmful.
On the other hand, $(F\tilde{F})_{\rm hidden}$
can induce $V_{\rm HE}\gg f_\pi^2m_\pi^2$, so the $U(1)_{PQ}$
breaking by $(F\tilde{F})_{\rm hidden}$ should be avoided
by choosing $\kappa_h=0$.
In fact, it is a plausible assumption that
the CS couplings 
$\kappa_{h,s,w}$ are quantized, and then 
the condition $\kappa_h=0$ does not correspond to an
unnatural fine tuning of parameters.
Another possibility is that
there are several $Z_2$-odd $U(1)$ gauge fields,
$C_M$ and $C^\pr_M$ for instance, having linearly independnet
CS couplings, and then there exists a combination $\tilde{C}_5
=C_5\cos\beta+C^\pr_5\sin\beta$ which can be identified as
the QCD axion since it does not couple to $(F\tilde{F})_{\rm hidden}$,
but couples to $(F\tilde{F})_{QCD}$.

\medskip

So far, we have noticed that all unwanted $U(1)_{PQ}$ breakings
in $S_{5D}$ can be suppressed enough as a consequence of
the 5D gauge symmetry $U(1)_C$ and the 5D locality.
For this,  it is required that all bulk matter fields
are neutral under $U(1)_C$
and the CS coupling of $C_5$ to $(F\tilde{F})_{\rm hidden}$
vanishes.
These conditions are stable against quantum corrections
including (nonperturbative)
quantum gravity effects, and thus is 
a natural framework.
Since the fifth dimension is compact, there can be nonzero
nonlocal effects breaking $U(1)_{PQ}$. However 
if all 5D matter fields $\Phi$
with masses $M_\Phi\lesssim \Lambda$ are neutral under $U(1)_C$,
such nonlocal effects are suppressed by
$e^{-\pi R\Lambda}$ where $R$ is the orbifold radius
and $\Lambda$ is 
a cutoff scale of our 5D orbifold field theory
\cite{pilo}. 
As a result, in our case
$V_{\rm HE}$ induced by nonlocal effects can be easily made
to satisfy the bound (\ref{heaxion}).
We thus conclude that the axion solution to the strong
CP problem can be realized naturally in 5D orbifold field
theory if the axion originates from a $Z_2$-odd 5D gauge field.

\medskip

Most of low energy axion physics is determined by its decay constant
$f_a$ \cite{strongcp}.
For a QCD axion solving the strong CP problem, cosmological and 
astrophysical arguments suggest $f_a=10^{10}\sim 10^{12}$
GeV which is lower than
$M_{Pl}$ 
by many orders of magnitudes.
In fact, for the axion originating from 
5D gauge field, a hierarchically small
$f_a/M_{Pl}$ is automatically obtained if $S^1/Z_2$ is {\it warped}.
To see this, let us consider a warped geometry with
$ds^2=G_{MN}dx^Mdx^N=e^{-2ky}\eta_{\mu\nu}dx^\mu dx^\nu+dy^2$
where $k$ denotes the AdS curvature \cite{rs}.
We first note that our previous arguments on the suppression
of unwanted $U(1)_{PQ}$ breakings are not affected by
warping.
To compute $f_a$, we start from the 5D action
\bea
S_{5D}&=&\int d^4x\,dy\, \sqrt{-G}\,\l(
\f{1}{4g_{5C}^2}C_{MN}C^{MN}
+\f{\kappa}{\sqrt{-G}}\epsilon^{MNPQR}C_MF^a_{NP}F^a_{QR}+...\r)\,,
\eea
where $F^a_{MN}$ denote the SM gauge field strengths.
To get the effective 4D action of $C_5$, one needs to integrate out $C_\mu$
by solving its equation of motion.
The equation of motion of $C_\mu$ under a $y$-independent
background of $C_5$ is given by
$\partial_5\l(e^{-2ky}C_{\mu 5}\r)=0$,
yielding
\bea
C_{\mu 5}=
\partial_\mu C_5-\partial_5C_\mu=\f{2\pi kR e^{2ky}}{e^{2\pi kR}-1}\,
\partial_\mu C_5.
\eea
The resulting 4D effective action of $C_5$ is
\bea
S_{\rm axion}^{4D}&=&
\int d^4x \, \l[
\, \f{\pi^2 kR^2}{2g_{5C}^2}\f{1}{e^{2\pi kR}-1}\partial_\mu C_5\partial^\mu
C_5
+2\pi \kappa R C_5\l(F\tilde{F}\r)_{QCD}
\r]\,.
\eea
From this, we find that the decay constant of the canonically normalized
axion field is given by
\bea
\label{decayconstant}
f_a&=&
\f{1}{64\pi^2\kappa}\l(\f{k}{g_{5C}^2}\f{1}{e^{2\pi kR}-1}\r)^{1/2}
\nonumber \\
&=& \f{1}{64\pi^2\kappa}
\l(\f{e^{\pi kR}}{e^{2\pi kR}-1}\r)\l(\f{k}{g^{}_{5C}M_5^{3/2}}\r)\,M_{Pl}\,,
\eea
where $M_5$ is the 5D Planck scale and $M^2_{Pl}=
M_5^3(1-e^{-2\pi kR})/k$ \cite{rs}.
Simple dimensional analysis suggests
$g_{5C}^2={\cal O}(1/M_5)$ and the CS coupling $\kappa={\cal O}(1)$.
Then the axion decay constant of $C_5$
in warped geometry  with $k\approx M_5$  is estimated to be
\bea
f_a={\cal O}\l(\f{e^{-\pi kR}M_{Pl}}{64 \pi^2}\r)\,,
\eea
so an intermediate axion scale $f_a=10^{10}-10^{12}$ GeV
is automatically obtained when $e^{-\pi kR}=10^{-4}-10^{-6}$.
However in this case we need $N=1$ supersymmetry to stabilize
the weak to axion scale hierarchy
$M_W/f_a=10^{-8}-10^{-10}$.
Note that if the standard model gauge fields propagate in bulk as
in our case, gauge couplings run logarithmically 
up to energy scales of order $k\approx M_5$,
so the SM gauge couplings can be unified 
at $M_{GUT}={\cal O}(M_5)\gg f_a$ \cite{unification}.

\medskip
To conclude, a QCD axion solving the strong CP problem
can arise naturally from a parity-odd gauge field $C_M$ in 
5D orbifold field theory.
All unwanted $U(1)_{PQ}$ breakings are suppressed enough by
the 5D gage symmetry of $C_M$ and the 5D locality, while
the required $U(1)_{PQ}$ breaking by the QCD anomaly originate from
gauge invariant local Chern-Simons coupling of $C_M$.
If the fifth dimension is warped,
the axion decay constant of $C_5$ is 
suppressed by an exponentially small warp factor
compared to $M_{Pl}$, while the scale of gauge unification
remains to be close to $M_{Pl}$.
Then the model
can have an intermediate axion scale
$f_a=10^{10}\sim 10^{12}$ GeV together with $M_{GUT}\gg f_a$
in a natural way.

\medskip

\bigskip

{\bf Acknowledgements}

\medskip

This work is supported by KRF PBRG 2002-070-C00022.

\end{document}